\documentclass[twocolumn,prl,superscriptaddress]{revtex4}
\usepackage{graphicx}
\usepackage{times}
\usepackage{amsmath}

\begin{document}

\title{Breakdown of a topological phase: 
Quantum phase transition in a loop gas model with tension}

\author{Simon Trebst}
\affiliation{Microsoft Research, Station Q, University of California,
Santa Barbara, CA 93106}
\author{Philipp Werner}
\affiliation{Department of Physics, Columbia University, 538 West
120th Street, New York, NY 10027}
\author{Matthias Troyer}
\affiliation{Theoretische Physik, Eidgen\"ossische Technische
Hochschule Z\"urich, CH-8093 Z\"urich, Switzerland}
\author{Kirill Shtengel}
\affiliation{Department of Physics and Astronomy, University of
California, Riverside, CA 92521}
\author{Chetan Nayak}
\affiliation{Microsoft Research, Station Q, University of California,
Santa Barbara, CA 93106}
\affiliation{Department of Physics and Astronomy, University of
California, Los Angeles, CA 90095}
\date{\today}

\begin{abstract}
  We study the stability of topological order against local
  perturbations by considering the effect of a magnetic field on a
  spin model -- the toric code -- which is in a topological phase.
  The model can be mapped onto a quantum loop gas where the
  perturbation introduces a bare loop tension.  When the loop tension
  is small, the topological order survives.  When it is large, it
  drives a continuous quantum phase transition into a magnetic state.
  The transition can be understood as the condensation of `magnetic'
  vortices, leading to confinement of the elementary `charge'
  excitations.  We also show how the topological order breaks
  down when the system is coupled to an Ohmic heat bath and discuss
  our results in the context of quantum computation applications.
 \end{abstract}

\maketitle

% Introduction ---------------------------------------------------------------------

% paragraph{Introduction --}
Topological phases are among the most remarkable phenomena in nature.
Although the underlying interactions between electrons in a solid are
not topologically invariant, their low-energy properties are. This
enhanced symmetry makes such phases an attractive platform for quantum
computation since it isolates the low-energy degrees of freedom from
local perturbations
-- a usual cause of errors \cite{ToricCode}. Tractable theoretical
models
with topological phases in frustrated magnets
\cite{ToricCode,magnets}, Josephson junction arrays
\cite{Ioffe02,Motrunich02}, or
cold atoms in traps \cite{ColdAtoms} 
have been proposed.  However, such phases have not, thus far, been
seen experimentally outside of the quantum Hall regime.
Is it because topological phases are very rare and
these models are adiabatically connected only to very small 
regions of the phase diagrams of real experimental systems?

In this paper, we take a first step towards answering this question.
We begin with the simplest exactly soluble model of a topological
phase \cite{ToricCode}, whose Hamiltonian is given below in
Eq.~(\ref{Eq:ToricCode}).  This model describes a frustrated magnet
with four-spin interactions similar to cyclic ring exchanges.  It is
closely related to the quantum dimer model \cite{DimerModel} for
frustrated magnets, which can be realized in Josephson junction arrays
\cite{Ioffe02}.  We consider perturbations of the soluble model that,
when sufficiently large, drive the system out of the topological
phase.
The question is, how large? A small answer would imply that this
topological phase is delicate and occupies a small portion of the
phase diagram.  This might explain the paucity of
experimentally-observed topological phases. Instead, we find that
`sufficiently large' is a magnetic field $h$ of order one ($h_c
\approx 0.6$)
in units of the basic four-spin plaquette interaction. 
Our numerical simulations demonstrate, for the first time, several key
signatures of the phase transition out of the topological phase,
including the finite-size degeneracy splitting of the topological
sectors, the condensation of `magnetic' excitations, and the
confinement of `electric' charges.

We also consider perturbing the system by coupling it to an 
Ohmic heat bath. When coupled to such a bath, a quantum mechanical
degree of freedom can undergo a transition
from coherent to incoherent behavior \cite{Leggett87}.
Recently, the effects of such a coupling on quantum phase transitions,
at which divergent numbers of quantum mechanical
degrees of freedom interact, have also been
studied \cite{Werner04}. In both contexts, the coupling to
the heat bath tends to make the system more
classical. Coherent quantum oscillations are suppressed, while
broken symmetry phases -- which are essentially
classical -- are stabilized. A topological phase is
quantum mechanical in nature.
We find that coupling the heat bath to the kinetic energy, i.e.
plaquette flip operator, does not destroy such a phase. 
However, when the dissipation is strong, the gap becomes very small,
and the topological phase may be too delicate to observe or use at
reasonable temperatures.  On the other hand, if the heat bath is
coupled to the classical state of each plaquette, the topological
phase is destroyed through a Kosterlitz-Thouless transition at a
dissipation strength of order one.

These results can be recast in quantum information language: the
ground states in different topological sectors are the different basis
states of an encoded quantum memory.  Quasiparticle excitations are
states outside of the code subspace.  The stability of the topological
phase, as measured by an energy gap $\Delta$ within a topological
sector (essentially the energy cost for a pair of quasiparticles),
translates into an error rate for topological qubits. At zero
temperature, errors are due to the virtual excitation of pairs of
quasiparticles, assuming that the system is shielded from
perturbations at frequencies higher than $\Delta$.  Such virtual
processes lead to a splitting between topological sectors
$\delta\! E \sim e^{-\Delta L/v}$, where $L$
is the system size and $v$ is a characteristic velocity. As the
temperature is increased, the thermal excitation of particles
eventually becomes more important and the error rate is $\sim
e^{-\beta\Delta}$ \footnote{In a special case of a topological phase
considered in \cite{Chamon05}, thermal activation \emph{always} dominates
quantum tunneling.}. (The actual concentration of excitations leading to
unrecoverable errors was studied in \cite{Wang03}.)

%- The toric code  ---------------------------------------

\begin{figure}[b]
  \includegraphics[height=25mm]{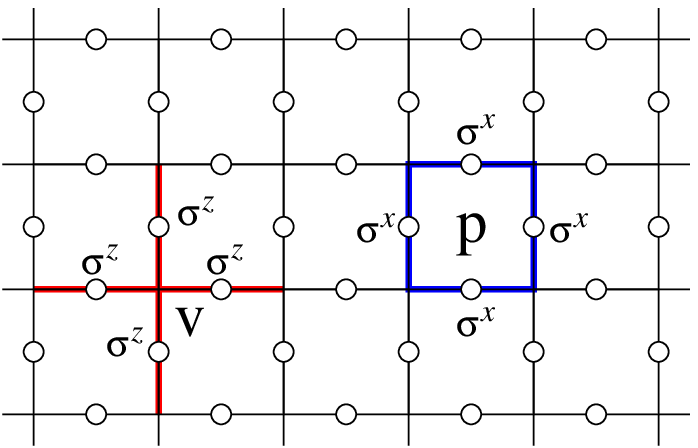}
  \hspace{3mm}
  \includegraphics[height=25mm]{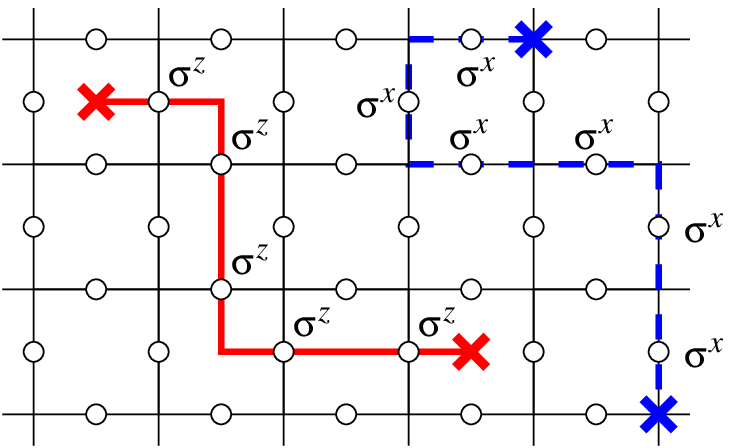}
  \caption{
    Left: Illustration of the toric code Hamiltonian (\ref{Eq:ToricCode}).
    Right: The elementary excitations above the loop gas ground state 
    are pairs of `magnetic vortices' on plaquettes connected by a string 
    of $\sigma^z$-operators (solid line) and `electric charges' on the 
    vertices connected by a string of $\sigma^x$-operators (dashed line).
  }
  \label{Fig:ToricCode}
\end{figure}

\paragraph{The model --}
We start with the toric code Hamiltonian \cite{ToricCode} 
\begin{equation}
  H_{\rm TC} = - A \sum_v \prod_{j\in\mbox{\scriptsize vertex}(v)}\sigma^z_j 
         - B  \sum_p \prod_{j\in\mbox{\scriptsize plaquette}(p)}\sigma^x_j \,,
  \label{Eq:ToricCode}
\end{equation}
where the $\sigma_i$ are $S=1/2$ quantum spins on the $2N$ edges of a
square lattice with $N$ vertices on a
torus. 
Since all terms in Hamiltonian (\ref{Eq:ToricCode}) commute with each
other, the model can be solved exactly \cite{ToricCode}. The
ground-state manifold can be described as a quantum loop gas where the
loops consist of chains of up-pointing $\sigma^z$-spins and the loop
fugacity is $d=1$.  On the torus there are four degenerate ground
states which can be classified by a winding number parity $P_{y/x} =
\prod_{i\in c_{x/y}} \sigma_i^z$ along a cut $c_{x/y}$ in the $x-$ or
$y-$direction.

% Loop tension & mapping
% -------------------------------------------------------------

Here we study the effect of perturbing the Hamiltonian
(\ref{Eq:ToricCode}) with a loop tension which can be introduced
either by a longitudinal magnetic field or local Ising interaction of
the form
\begin{equation}
  H = H_{\rm TC} - h \sum_i \sigma_i^z - J \sum_{\langle ij \rangle} 
\sigma_i^z\sigma_j^z \,,
  \label{Eq:LoopTension}
\end{equation}
where $h$ ($J$) is the strength of the magnetic field (Ising
interaction). These are the dominant perturbations expected in a
physical implementation, e.g. in a Josephson junction implementation
\cite{Ioffe02,Motrunich02} they arise from electric potential
perturbations or Coulomb interactions between neighboring 
quantum dots. 
We discuss this model in the limit of a large charge gap,
i.e. $A\gg B,h,J$, where it becomes equivalent to the `even' Ising
gauge theory \cite{IGT}.  The low-energy sector has no free charges
and any state is described by a collection of loops that can be
obtained from a reference state (e.g. all $\sigma_i^z={1}/{2}$) by a
sequence of plaquette flips.  Let us introduce a new plaquette spin
operator $\mu_p$ with eigenvalues $\mu_p^z = (-1)^{n_p}/2$ where $n_p$
is the number of times a given plaquette $p$ has been flipped,
counting from the reference state.  Then $\sigma_i^z = 2 \mu_p^z
\mu_q^z$, where $p$ and $q$ are the plaquettes separated by the edge
$i$. The plaquette flip term in Eq.~(\ref{Eq:ToricCode}) becomes
$-4B\sum_{p} \mu_p^x$. In the new variables, Hamiltonian
(\ref{Eq:LoopTension}) becomes equivalent to the transverse field
Ising model (with both nearest and next-nearest neighbor Ising
interactions) in a basis restricted to loop states. Independent of
the choice of basis states the system 
orders at a critical field strength $(h/B)_c = 0.65695(2)$ determined
by continuous-time quantum Monte Carlo simulations \cite{TIM}.  The
transverse field Ising model for the plaquette spins can be mapped to
a classical (2+1)-dimensional Ising model:
\begin{equation}
  \mathcal{H}_{\text{cl}} 
  = - K_{\tau} \sum_{\tau, p} S_p(\tau)
S_p(\tau+\Delta\tau)- K \sum_{\tau, \langle p,q \rangle}S_p(\tau) S_q(\tau), 
  \label{Eq:3DIsing}
\end{equation}
where the imaginary time $\tau$ is discretized in steps of
$\Delta\tau$ and $S_p \equiv 2 \mu_p^z = \pm 1$.  The magnetic field
$h$ and the local Ising interaction $J$ introduce a nearest and
next-nearest neighbor interaction between the classical spins
$S_p(\tau)$, respectively.
Since the exact nature of this Ising interaction does not play a role
in the following we do not discuss the case of next-nearest neighbor
couplings in detail.  The coupling along real-space coordinates is
then given by $K = \frac{1}{2} \Delta\tau \cdot h$ and along imaginary
time by $K_{\tau} = -\frac{1}{2} \ln \left[ \tanh (\Delta\tau \cdot B)
\right]$.  The model (\ref{Eq:3DIsing}) describes the well-known
continuous magnetic phase transition of the 3D Ising model. For
isotropic interactions, $K=K_{\tau}$, the critical coupling has been
determined with high precision to be $K_c = 0.221 659 5(26)$
\cite{Ferrenberg:91}.  Setting $B=1$ this gives a critical loop
tension $h_c = 0.582 24$ with isotropic couplings at $\Delta \tau =
0.761 403$.
\begin{figure}[t]
  \includegraphics[height=58mm]{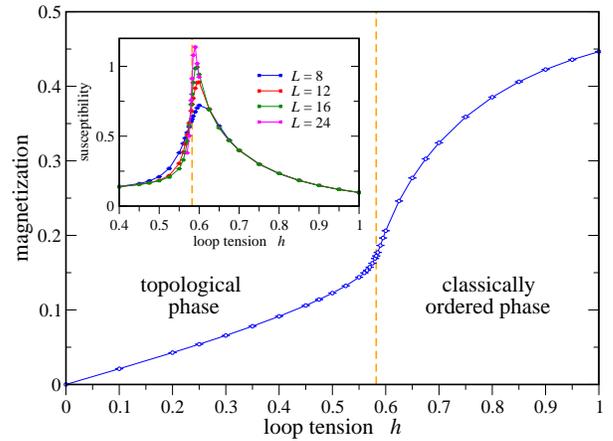}
  \caption{
    (Color online) Magnetization $M = \langle \sum_i \sigma_i^z
    \rangle/2N$ versus loop tension (magnetic field).  The topological
    phase survives for small loop tension where an almost constant
    susceptibility (see inset) leads to a linear increase of $M$.
    Above the critical loop tension (dashed line) the system
    approaches the fully polarized state.  }
  \label{Fig:Magnetization}
\end{figure}
% Phase transition for loop tension ---------------------------------------
\begin{figure}[b]
  \includegraphics[width=86mm]{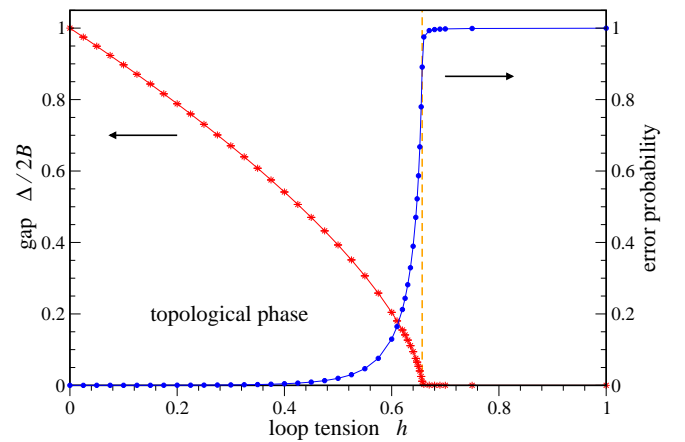}
  \caption{
    (Color online) Excitation gap for magnetic vortex excitations
    (star symbols) versus the loop tension.
    At the critical loop tension the gap closes and the magnetic
    vortices condense.  Right ordinate: Rate $\exp(-\beta
    \Delta)$ for tunneling processes between topological sectors
    (filled circles) for $\beta=10$.  }
  \label{Fig:TIM}
\end{figure}

The magnetic susceptibility diverges at the transition and the
magnetization $\langle \sum_i \sigma_i^z \rangle/2N$ shows a corresponding
kink, as shown in Fig.~\ref{Fig:Magnetization}. Although
this is not a symmetry-breaking transition, the
analogous transition driven by next-nearest interaction $J$ is a
{\em continuous} quantum phase transition from a topologically 
ordered quantum state to a broken symmetry state \cite{ContinuousPhaseTransition}.
The magnetic transition can also be understood in terms of the condensation
of `magnetic vortices', plaquettes with $\prod_{j}\sigma^x_j=-1$. 
While for the original Hamiltonian (\ref{Eq:ToricCode}), the gap to these
excitations is $\Delta = 2B$, they become gapless and condense at the critical
loop tension, as shown in Fig.~\ref{Fig:TIM}. 
The gap has been estimated from measurements of the imaginary time
correlation length $\xi_\tau$ as $\Delta \propto 1/\xi_\tau$ which we have
calculated applying continuous-time quantum Monte Carlo simulations using the
ALPS looper code \cite{ALPS,LooperCode}). The thermal excitation of pairs
of magnetic vortices occurs with probability $\sim \exp(-\beta\Delta)$,
also shown in Fig.~\ref{Fig:TIM}.
\begin{figure}[t]
  \includegraphics[height=58mm]{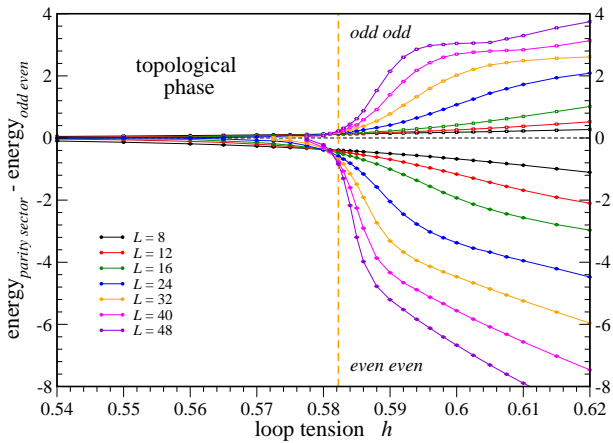}
  \caption{
    (Color online)
    Splitting of the energy degeneracy
    for the four topological sectors defined by even or odd winding
    number parities along the $x$- and $y$-directions.  The
    intermediate \{even-odd / odd-even\}-parity sector was taken as a
    reference (dashed line).  }
  \label{Fig:Energy}
\end{figure}

\begin{figure}[t]
  \includegraphics[height=58mm]{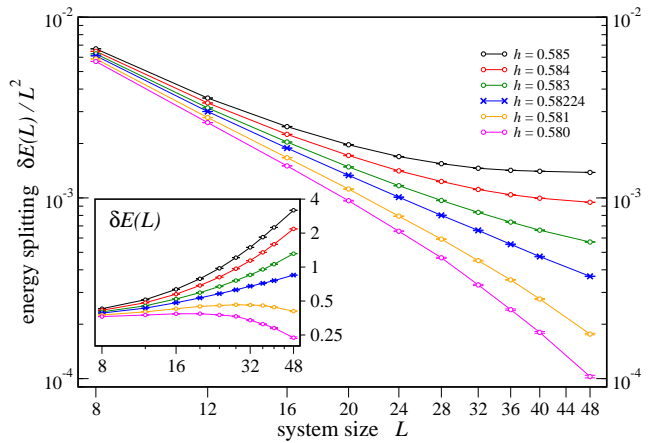}
  \caption{
    (Color online) Finite-size scaling of the energy splitting between
    the topological sectors around the quantum phase transition for
    $\beta = 10 L$.  At the critical loop tension (crosses) we find
    power-law scaling $\delta\! E(L) \propto L^{2-z}$ with exponent $z
    = 1.42 \pm 0.02$.  }
  \label{Fig:EnergyScaling}
  \vspace{-0.75mm}
\end{figure}

\paragraph{Topological order --} 
The breakdown of topological order at
the phase transition can be seen from the energy
splitting $\delta\! E$ between the ground-states for the various
topological sectors. When winding 
parities are used as basis states for a quantum memory, this splitting
causes phase errors.  (The absence of `electric' charges precludes any
transitions between different winding parities so bit flip errors cannot
occur.)  In the topological phase, the virtual excitation of quasiparticles
leads to a small splitting $\delta\! E \propto \exp(-\Delta L/v)$ between the
topological sectors.  In the classically-ordered phase, on the other
hand, the energy splitting should scale with $L$, which corresponds to
the energy cost of a loop in the ordered ground state.  As the winding
parity is conserved by imaginary time spin-flip operations, we
can simulate the system in one of the topological sectors by defining
an initial spin configuration that corresponds to the respective limit
for large loop tension. 
Fig.~\ref{Fig:Energy} shows the calculated splitting
for various system sizes in the vicinity of the critical loop tension.
At the phase transition, the behavior qualitatively changes from
power-law scaling for strong loop tension to an exponential
suppression in the topological phase for small loop tension as
discussed above. A more quantitative picture arises from the
finite-size scaling analysis of the energy splitting $\delta\!  E(L)$
between the \{even-odd\}- and \{even-even\}-parity sectors shown in
Fig.~\ref{Fig:EnergyScaling}.  For the critical loop tension we find a
power-law scaling $\delta\!  E(L) \propto L^{2-z}$ with an exponent $z
= 1.42 \pm 0.02$.
Below the critical value the scaling 
turns into exponential scaling as expected for the topological phase.

% Confinement transition for "electric charges" ---------------------------------------

\begin{figure}[t]
  \includegraphics[height=58mm]{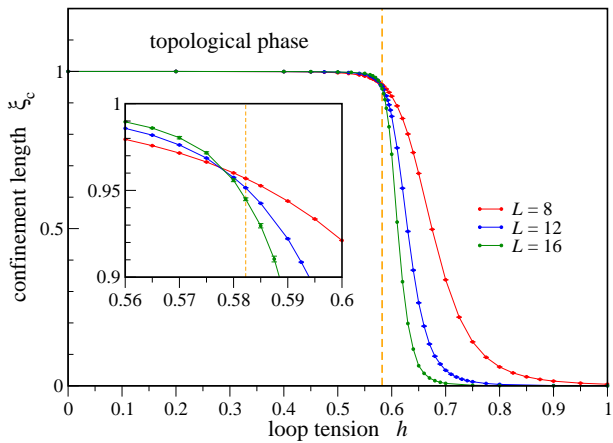}
  \caption{
    (Color online) Confinement length of two electric charge
excitations versus loop tension. The confinement transition occurs at
the same critical loop tension (dashed line) as the magnetic
transition.
  }
  \label{Fig:Confinement}
\end{figure}

\paragraph{Confinement transition --} For the loop gas Hamiltonian
($\ref{Eq:ToricCode}$) the elementary electric charge excitations (end-points
of an open loop) are deconfined. For strong
loop tension, however these excitations are expected to become
confined, thereby eliminating all open loops. We can study this
confinement transition in our simulations of model (\ref{Eq:3DIsing})
by artificially introducing pairs of electric charge excitations for
the sampled loop configurations. This allows us to measure the
confinement length $\xi_c$ as the square root of the average second
moment of the distance between the two excitations, which for a torus
with even extent $L$ has to be normalized by a factor $6/(L^2+2)$.
The measured confinement length $\xi_c$ shown in
Fig.~\ref{Fig:Confinement} clearly demonstrates that electric charges
remain deconfined for the full extent of the topological phase and
that the confinement transitions occurs simultaneously with the
magnetic phase transition. At the critical loop tension the
confinement lengths $\xi_c(L)$ for various system sizes cross which
demonstrates that the confinement length diverges with the \emph{same}
critical exponents as the magnetic correlation length $\xi$ and there
is only one length scale describing the phase transition.  This can be
seen from the following finite-size scaling argument: We may write the
finite-size scaling behavior for the confinement length as
$\xi_c(\tau, L) \propto \tau^{-\nu} f(L \tau^{\chi})$ where $\nu$
($\chi$) is the critical exponent of the correlation (confinement)
length respectively and $\tau = h-h_c$.  With
$\tilde{f}(L^{1/\chi}\tau) = (L^{1/\chi}\tau)^{-\nu} f(L \tau^\chi)$
we obtain for the confinement length $\xi_c(\tau, L) \propto L^{\nu /
  \chi} \tilde{f} (L^{1/\chi} \tau)$.  At the phase transition the
existence of a crossing point $\xi_c(0, L)/L \propto L^{\nu / \chi -
  1} = {\rm const}$ then implies $\chi = \nu$.  For our model without
dynamical electric charges, this measure of the confinement of test
charges is closely related to the calculation of a Wilson loop
expectation value.
In the presence of dynamical electric charges, this becomes trickier;
Polyakov loops have been used as an order parameter
for the {\em finite} temperature transition of the
3D Ising gauge model\cite{LGT}.

 % Dissipation -------------------------------------------------------------

\paragraph{Dissipation --}
Finally, we discuss the effect of dissipation when Hamiltonian
(\ref{Eq:ToricCode}) is coupled to an Ohmic heat bath. Since our model
excludes dynamical electric charges, we do not consider coupling a
heat bath directly to $\sigma^x_i$.
Instead, we first examine coupling a heat bath to $\mu^x_p$ so that a 
`phonon' is created when a plaquette flips.
This type of dissipation could occur in a Josephson junction model
\cite{Schoen:90} or in a spin model through the spin-phonon coupling.
The standard procedure \cite{Caldeira} for a linear spectral density
(`Ohmic' dissipation) results in an effective action
for independent Ising chains with long-range couplings in an external
\emph{longitudinal} magnetic field (which here means parallel to
$\mu^x$).  We note that as a consequence of the Lee--Yang theorem
\cite{Lee:52}, there can be \emph{no singularities} of the respective
partition function at any real non-zero 
longitudinal field, ruling out the existence of a quantum phase
transition for this model. In particular, this implies that the magnetic gap
remains finite for any dissipation strength!

An entirely different behavior arises if dissipation is coupled such
that it stabilizes the `classical' state of the system.  Coupling the
bath to either $\sigma^z_i$ or ${\mu^z_p}$ stabilizes the classical
state of a single spin or a plaquette, respectively. We consider the
latter as it should be more effective at damping quantum fluctuations,
although we expect the former to have similar physics.
The same procedure as above then leads to a model for
decoupled Ising chains given by
\begin{multline}
\mathcal{H}_{\text{cl}} 
= - K_{\tau} \sum_{\tau, p} S_p(\tau)
S_p(\tau+\Delta\tau)
\\
-\frac{\alpha}{2}\sum_{\tau<\tau',p}
\Big(\frac{\pi}{N_\tau}\Big)^2\frac{S_p(\tau)S_p(\tau')}{\sin^{2}(\frac{
\pi}{N_\tau}|\tau-\tau'|)} \,,
\label{eq:dissip_action_z}
\end{multline}
where the parameter $\alpha$ measures the dissipation strength.
This model has been well studied \cite{Dissipation} and is known to
exhibit a Thouless-type phase transition into a classically
disordered, fluctuationless phase.  The critical value $\alpha_c$ of
this transition depends weakly on the cutoff in the long-range
interaction; in our simulations $\alpha_c\approx 0.7$.
At the transition the magnetic gap vanishes, in sharp contrast to the
previous case.

\begin{figure}[t]
  \includegraphics[height=58mm]{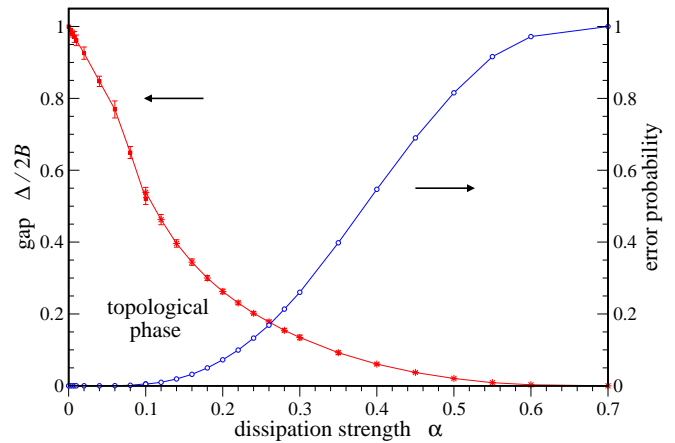}
  \caption{
    (Color online) Gap and error probability versus dissipation
    strength estimated from the correlation time $\xi_\tau$ of the
    dissipative Ising chain (\ref{eq:dissip_action_z}).
   }
  \label{Fig:Dissipation}
  \vspace{-0.5mm}
\end{figure}

Due to the long range interactions introduced by the dissipative
coupling, the spin-spin correlations will asymptotically decay as
$1/\tau^2$ \cite{Griffiths:67}. It is therefore a non-trivial task to
define a correlation-time and hence to estimate the excitation gap.
For $\alpha\lesssim 0.1$ one observes an exponential decay of the
correlation function onto the asymptotic $1/\tau^2$ behavior, and
subtracting the $1/\tau^2$-contribution allows one to estimate
$\xi_\tau$, which is found to grow linearly with $\alpha$. For
$\alpha>0.1$, this procedure can no longer be used. In this region we
estimate $\xi_\tau$ from the asymptotic decay of the correlations,
proportional to $(\xi_\tau/\tau)^2$. These $\xi_\tau$s grow
approximately exponentially in the region $0.1\lesssim \alpha
\lesssim \alpha_c$. Alternatively, one could define a correlation-time
from the crossover scale where the short-time behavior crosses over to
the asymptotic $1/\tau^2$ form. This definition in the spirit of a
`Josephson length' \cite{Chakravarty:95} yields (up to a normalization
factor) the same results as the $\xi_\tau$ extracted from the
asymptotic decay.  The gap estimated from the inverse correlation
function, as well as the error probability is plotted in
Fig.~\ref{Fig:Dissipation}. We find that the error probability remains
negligibly small below the crossover value $\alpha\approx 0.1$.

% Summary / Conclusions ----------------------------------------------------

\paragraph{Outlook --} 
We have shown that the topological phase which governs the
toric code model \cite{ToricCode} actually exists in an extended
region of phase space around the soluble point. It is stable against
deviations of the system Hamiltonian from the ideal one and also the
coupling of the system to its environment.  In general, this
demonstrates that a system does not necessarily have to be
particularly fine-tuned to reach a topological phase. The paucity of
their experimental observations may thus be due not to some intrinsic
delicateness of such phases, but rather to the experimental subtlety
involved in identifying them. In future work, these conclusions need
to be tested in other, more exotic topological phases which support universal
topological quantum computation \cite{Freedman01}.

% Acknowledgments ---------------------------------------------------------

We thank E. Ardonne, L. Balents, S. Chakravarty, and A. Kitaev for
stimulating discussions.

% References ---------------------------------------------------------

\end{document}